\documentclass[aps,prb,twocolumn,showpacs,amsmath,amssymb,10pt,letterpaper,floatfix]{revtex4-1}
      \newcommand{\beq}{\begin{equation}}
      \newcommand{\eeq}{\end{equation}}
      \newcommand{\beqa}{\begin{eqnarray}}
      \newcommand{\eeqa}{\end{eqnarray}}
      
      \newcommand{\nn}{\nonumber}

      \newcommand{\bra}{\left\langle}
      \newcommand{\ket}{\right\rangle}

       \newcommand{\del}{\partial}

      \newcommand{\be}{\beta}
      \newcommand{\ga}{\gamma}
      
      \newcommand{\ep}{\epsilon}
      
      \newcommand{\la}{\lambda}

    \renewcommand{\(}{\left(}
    \renewcommand{\)}{\right)}
\usepackage{graphicx}% Include figure files
\usepackage{dcolumn}% Align table columns on decimal point
\usepackage{bm}% bold math
\usepackage{hyperref}
\usepackage{amsmath}
\begin{document}
\title
{The generating function for the connected correlator in the 
random energy model and its effective potential
    }
 
 \author{Hisamitsu Mukaida}
 \email{mukaida@saitama-med.ac.jp}
 \affiliation{Department of Physics, Saitama Medical University, 
 38 Moro-Hongo, Iruma-gun, Saitama, 350-0495, Japan
 }
 \date{\today}

\begin{abstract}
Starting with two copies of the random energy model coupled with independent 
magnetic fields, the generating function for the connected correlator of
the magnetization is exactly derived. 
Without use of the replica trick, it is shown that the Hessian of  the generating function
is symmetric under exchanging the two copies when the system is finite, 
but  the symmetry is spontaneously broken in the low-temperature phase.   
It can be regarded as a 
rigorous realization of  the replica symmetry breaking.  
 The corresponding effective potential, which has two independent 
 variables conjugate to the magnetic fields,  
is also calculated. It is singular when the two variables coincide. The singularity 
is consistent with that observed in the effective potentials of  
short-ranged disordered systems 
in the context of the functional renormalization group. 
\end{abstract}

 %75.10.Hk	Classical spin models
 %75.10.Nr	Spin-glass and other random models 
 %64.60.ae	Renormalization-group theory

\maketitle

\section{Introduction}
A technical difficulty in theoretical study of quenched disordered systems 
originates from inhomogeneity due to disordered environment.  
In those systems, we first take the thermal average of physical quantities in a fixed disordered environment and 
then we need to take the average over the disorder. However, if we can first average out  
the disorder, problems in those systems
 will be more tractable.   Several methods to make it possible
 are developed in the last four decades and they expose peculiarities in 
 quenched disordered systems. 
 
One of the most popular method will be the replica trick \cite{MPV87}, 
where identical $n$ copies (replicas) of the 
system are introduced. In mean-field models such as 
the Sherrington-Kirkpatrick model \cite{SK75}
 or the random energy model (REM) \cite{D80,D81}, glassy behavior is  
revealed together with the replica symmetry breaking (RSB).  In order to show the RSB, 
the limit $n \to 0$ is taken despite that $n$ is a natural number.  The RSB is 
a peculiar feature of quenched disordered systems in the sense that
 we do not know a mathematical reason why it gives the correct answer \cite{D11}. 

Another peculiarity we treat here is that a non-analytic potential appearing as a fixed point 
of a flow equation of the 
functional renormalization group (FRG) 
in short-ranged disorder models \cite{F86,Fe01,LW02,TT04,WL07,LMW08,TT08}.
  Because of
 a semi-quantitative 
argument why the non-analytic fixed point appears and of 
consistency with other methods \cite{WL07}, 
existence of the non-analyticity in the fixed-point potential 
is quite plausible. 
However, we do not know its robustness against 
higher-order corrections to the flow equation.  

Therefore it is worthwhile to clearly show  
existence of these peculiarities in disordered systems 
by solving a simple model exactly. 
In this paper, dealing with the REM, we derive the 
exact generating function for the connected two-point function  of the 
magnetization. In this case, as pointed out in the literature \cite{LW02, TT04, LMW08, TT08}, 
we need to introduce two copies of the REM coupled with independent external sources 
and to take the average over disorder.  Here, we do not use 
the replica trick for mathematical justification, but use 
the normalized partition function. 
The validity of the normalization in quenched disordered systems 
is emphasized in Refs.~\onlinecite{WL07, KL09},  and is realized by the 
Keldysh formalism \cite{CLN99} or by 
the supersymmeric method \cite{W05}.  In the REM, it is simply carried out 
using an integral representation of the normalization factor. 

The generating function obtained in this way is symmetric under the exchanging the two external sources. 
However, computing the Hessian of this function, 
we can show  in the low-temperature phase that the symmetry is  spontaneously broken 
in the usual sense of the statistical machanics.
That can be a mathematically well-defined counterpart of the usual RSB.   
Furthermore, we show that 
the effective potential, which is obtained by the Legendre transformation of the 
generating function, becomes
 non-analytic in accordance with the symmetry breaking. 
 It is quite similar to the result in study of
  random manifolds employing the functional renormalization group \cite{LW02, LMW08}. 

This paper is organized as follows: 
in the next section, we recall the definition of the REM in a uniform magnetic field 
and define the generating function for connected correlation functions 
of the total magnetization.  We also review that 
$n$ copies of the system independently coupled with magnetic fields 
are needed for deriving the $n$-point correlation function \cite{LW02, TT04, LMW08, TT08}.
In section \ref{sec_gen}, 
we calculate the exact generating function for the connected two-point function 
when the system is finite. We see that it has 
a symmetry exchanging two copies of the system.  
As a result, the Hessian of the generating function 
 becomes a replica symmetric matrix at the zero-magnetic field. 
In section \ref{sec_asy}, we study the asymptotic behavior of the generating function 
when the system approaches the thermodynamic limit.  
If the magnetic fields are turned off after the thermodynamic limit is taken, 
the Hessian is not replica symmetric anymore in the low-temperature phase. 
It means that the symmetry is spontaneously broken. 
In section \ref{sec_eff}, performing the Legendre transformation 
to the generating function in the thermodynamic limit, 
we obtain the exact effective potential.  Corresponding to the two external 
sources in the generating function, 
the effective potential has two independent variables, $\varphi_1$ and $\varphi_2$. 
We show that it is analytic in the high-temperature phase, 
while 
it is singular on  $\varphi_1 \varphi_2=0$ or on $\varphi_1 = \varphi_2$ 
in the low temperature phase. 
A physical interpretation of this singularity is also presented. 
The last section is devoted to summary and discussion.

\section{The REM in a magnetic field and its generating function}
\label{sec_REM}
The random energy model (REM) is defined on configurations of 
$N$ spins $\{\sigma_i\}$, $(i=1, ... ,N)$, 
each of which takes the values of $\pm 1$. When there is no external field, the energy $E$ of 
a spin configuration is completely independent of how the configuration is. 
It just follows a gaussian probability density $P(E)$ describing disorder environment.  
When a uniform magnetic field $H$ is turned on, 
the energy $E$ gets dependence on the magnetization $M := \sum_{i=1}^N \sigma_i$
and is modified to $E - HM$.  

For a precise description, it is 
convenient to  classify all the spin configurations by  values of $M$ \cite{DW01}.  
Since the number of the configurations with the magnetization $M$
is $n(M):=\binom{N}{(N+M)/2}$, we can label all the states 
with the two numbers $M$ and $k$,  where $M \in \{ -N, -N+2, ..., N-2, N\}$ and 
$k\in \{1, 2, ..., n(M)\}$. The energy of the state labelled by $(M, k)$ is $E_{M, k} - HM$. 
Here $E_{M, k}$ is a random variable obeying the following gaussian probability density 
independent of 
$M$ and $k$:
\beq
  P(E_{M,k}) := \frac{1}{\sqrt{\pi N J^2}} \exp \left(-\frac{E_{M, k}^2}{N J^2} \right), 
  \label{def_P}
\eeq
which defines the average over disorder. We denote it by the overline  as 
\beq
   \overline{X} := \int \prod_{M} \prod_{k=1}^{n(M)} P(E_{M, k}) \, X \, dE_{M, k}. 
\eeq
The partition function is given by 
\beq
  Z(H) := \sum_M \sum_{k=1}^{n(M)}e^{-\be E_{M, k} + \be M H}.  
  \label{def_Z}
\eeq
Note that $\sum_M n(M)$ equals the total number of configurations $2^N$. 
 
Let us obtain the generating function for the disorder average of the connected 
correlation functions of $M$. 
If we ignore the average over the disorder, 
 the generating function is given by $\be^{-1} \log Z(H)$. Taking the derivative 
with respect to $H$, we can obtain connected correlation functions of  
$M$.  
However, taking into account the random average, if we start with $\overline{Z(H)}$, 
we have to take care of a couple of things.  To see this, taking the derivative of 
$\log \overline{Z(H)}$, one finds that 
\beq
\be^{-1} \left. \frac{\del \log \overline{Z(H)}}{\del H} \right|_{H=0} = 
\frac{\overline{\bra M \ket Z(0)}}{\overline{Z(0)}}, 
 \label{form_pre-one}
\eeq
where the angle brackets denotes the thermal average with zero-magnetic field as follows:
\beq
  \bra X \ket := \frac{1}{Z(0)} \sum_{k=1}^{n(M)} X e^{-\be E_{M, k}}. 
\eeq
The result (\ref{form_pre-one}) differs from the one-point function. 
One of the usual ways of getting around the problem is the replica trick.  Namely, 
we use the partition function for $n$ ($n$ is a positive integer) copies of the model 
$\overline{Z(H)^n}$ instead of $\overline{Z(H)}$. Define the generating function  by 
$(\be n)^{-1}\log \overline{Z(H)^n}$. After taking the derivative, letting $n \to 0$, we 
formally obtain the correct one-point function.  However, taking the limit $n \to 0$ is not 
a procedure mathematically justified, so that we do not use it in the 
present paper.  Instead,
 we use the normalized partition function defined as 
\beq
  z(H) := \frac{Z(H)}{Z(0)}. 
  \label{def_z}
\eeq
Inserting $z(H)$ instead of $Z(H)$ in (\ref{form_pre-one}) 
and using the normalization condition $z(0) =1$, 
we can obtain the correct one-point function. 

However, it is not enough to obtain the correlation functions. 
In fact, taking the second derivative, we find that 
\beq
  \left. \be^{-2} \frac{\del^2 \log \overline{z(H)}}{\del H^2} \right|_{H=0}
   = \overline{\bra M^2\ket} - \overline{\bra M\ket} \  \overline{ \bra M\ket}, 
\eeq
which is not the disorder average of the connected two-point function.  In order to 
obtain the correct one,  we introduce  two copies of the system 
coupled with independent magnetic 
fields \cite{LW02, TT04, LMW08, TT08}.  Namely, we define 
\beq
  W(H_1, H_2) := \be^{-1} \log \overline{z(H_1) z(H_2)}. 
  \label{def_W}
\eeq
Then it is easily seen that 
\beqa
  \be^{-1} \left. \del_1^2 W(H_1, H_2) \right|_{H_1 = H_2= 0} 
  = \overline{\bra M^2\ket} - \overline{\bra M\ket} \  \overline{ \bra M\ket}
  \nn\\
 \be^{-1}  \left. \del_1 \del_2 W(H_1, H_2) \right|_{H_1 = H_2= 0} 
  = \overline{\bra M \ket^2} - \overline{\bra M\ket} \  \overline{ \bra M\ket}, 
\eeqa
where $\del_a$ $(a=1,2)$ means the derivative with respect to $H_a$. 
We obtain the connected two-point function from  the right-hand side of 
the following formula:
\begin{widetext}
\beq
  \overline{\bra M^2\ket - \bra M \ket^2} = 
    \be^{-1} \left. \left(\del_1^2 W(H_1, H_2) 
   -   \del_1 \del_2 W(H_1, H_2) \right)\right|_{H_1 = H_2= 0}. 
   \label{form_M;M}
\eeq
\end{widetext}
In general, if we want to generate the connected $n$-point function, 
we need the following generalization of (\ref{def_W}):
\beq
  W(H_1, ..., H_n) :=  \be^{-1} \log \overline{\prod_{k=1}^n  z(H_k)}. 
\eeq
For deriving the connected $m$-point function with $m<n$, 
we may just put $H_{m+1} = \cdots = H_{n}=0$.  Thus, 
$W(H_1, ..., H_n)$ contains all information up to the $n$-point functions. 
In this paper we investigate the simplest but nontrivial case,  
$n=2$. 

\section{The generating function in the finite system}
\label{sec_gen}
Now let us derive the generating function (\ref{def_W}).  
From the definition (\ref{def_z}),  we have 
\beq
  \overline{z(H_1) z(H_2)} = \overline{\left( \frac{Z(H_1) Z(H_2)}{Z(0)^2} \right)}. 
\eeq
The denominator makes the computation difficult. 
There are a couple of ways of ensuring the normalization condition such as 
the Schwinger-Keldysh approach \cite{CLN99, KL09} 
or the supersymmetric method \cite{W05}. 
In the REM, it can be established using the following representation:
\beq
  \frac{1}{Z(0)^2} = \int_0^\infty  \, t \, e^{-t Z(0)} dt. 
\eeq
Then we have 
\beq
  \overline{z(H_1) z(H_2)} = \int_0^\infty  dt \, t \ \overline{Z(H_1) Z(H_2) \, e^{-t Z(0)} }. 
  \label{form_z1z2-1}
\eeq 

The function $\overline{e^{-tZ(0)}}$ is first introduced by Derrida \cite{D81} and studied in detail. 
Namely, using the fact that the energies $\{E_{M, k} \}$ follow the probability density (\ref{def_P}) independently,  we can write  
\beqa
  \overline{e^{-t Z(0)}} &=& \overline{\prod_{M} \prod_{k=1}^{n(M)} e^{-t e^{-\be E_{M, k}}}} 
  = \prod_{M} \prod_{k=1}^{n(M)} \overline{e^{-t e^{-\be E_{M, k}}}} 
  \nn\\
  &=&  \left(f(t)\right)^{2^N}, 
  \label{form_e-tZ}
\eeqa
where
\beq
  f(t) :=  \overline{e^{-t e^{-\be E_{M, k}}}}. 
\label{def_ft}
\eeq
By definition,  it is immediately derived that 
\beqa
  f'(t) &=&   - \overline{e^{-\be E_{M, k} -t e^{-\be E_{M, k}}}} \nn\\
  f''(t) &=& \overline{e^{-2 \be E_{M, k} -t e^{-\be E_{M, k}}}}. 
  \label{form_f'f''}
\eeqa
Using (\ref{form_e-tZ}), (\ref{form_f'f''}) and (\ref{def_Z}) in the right-hand side of 
(\ref{form_z1z2-1}), we find that 
\begin{widetext}
\beqa
  \overline{z(H_1)z(H_2)} = 
  2^{-N}\sum_{M} e^{\be (H_1+H_2) M} \, n(M) \, J_N
%  \nn\\
%  && 
  + 2^{-2N}\sum_{M_1, M_2} e^{\be( H_1 M_1 + H_2 M_2)} \, n(M_1) n(M_2) I_N, 
  \label{form_z1z2}
\eeqa
\end{widetext}
where 
\beqa
  J_N &:=& 2^{N} \int_0^\infty dt \, t \, \left( f''(t) \( f(t) \)^{2^N-1} -  f'(t)^2 \( f(t) \)^{2^N-2} \right)
  \nn\\
  I_N &:=& 2^{2N}\int_0^\infty dt \, t \, f'(t)^2 \( f(t) \)^{2^N-2}. 
  \label{def_INJN}
\eeqa
Setting $H_1=H_2=0$ in (\ref{form_z1z2}),  we see the following relationship 
between $I_N$ and $J_N$: 
\beqa
  1 = \overline{z(0) z(0)} &=&  J_N + I_N. 
\eeqa
As a result, we can write 
\beq
 \overline{z(H_1) z(H_2)} = A+B
 \eeq
with 
\beqa 
  A &:=& 2^{-N} (1-I_N)\sum_{M} e^{\be (H_1+H_2) M} \, n(M)
 \nn\\
 B &:=&  2^{-2 N} I_N  \sum_{M_1, M_2} e^{\be (H_1 M_1 + H_2 M_2)} n(M_1) n(M_2).  
  \label{def_AB}
\eeqa
Thus, the generating function defined in (\ref{def_W}) is  written as  
$W(H_1, H_2) = \be^{-1}\log (A+B)$. 
Hereafter, we treat its density defined by 
\beq
  w_N(H_1, H_2) := \frac{1}{N} W(H_1, H_2) = \frac{1}{N \be}\log \( A+ B \)
  \label{def_wN}
\eeq
instead of $W(H_1, H_2)$ itself. 

From (\ref{def_AB}) and (\ref{def_wN}), it is obvious that $w_N(H_1, H_2)$ is 
symmetric under exchanging $H_1$ and $H_2$.  This yields the fact that 
 the coefficient of $H_1^k H_2^l$ in $w_N(H_1, H_2)$ must be the same 
 as that of $H_1^l H_2^k$ for arbitrary non-negative integers $k$ and $l$. 
It means that  
\beq
  \del_1^{k} \del_{2}^{l} w_N(0, 0) = \del_1^{l} \del_{2}^{k} w_N(0, 0). 
  \label{form_sym}
\eeq
In particular, the Hessian matrix  of $w_N(H_1, H_2)$ defined by 
\beq
w^{(2)}_N(H_1, H_2)_{ab} := \del_a \del_b w_N(H_1, H_2) 
\label{def_hesswN}
\eeq 
must be a replica symmetric matrix. 
We can explicitly derive it employing the formula
\beq
  \sum_M M n(M) = 0, \ \ 
  \sum_M M^2 n(M) = N 2^N. 
\eeq
The result is 
\beq
   w^{(2)}_N(0, 0) = \left(
 \begin{array}{cc}
 \be & \be(1-I_N)  \\ 
  \be(1-I_N)  & \be
  \end{array}
  \right) 
  \label{form_finiteN}
\eeq
 for an arbitrary finite $N$. 

Let us call the two copies of the REM considered here the copy 1 and the copy 2.
Suppose that they are respectively coupled with $H_1$ and $H_2$.  
Exchanging $H_1$ and $H_2$ means exchanging the two copies 1 and 2. 
Thus the above symmetry is similar to the usual replica symmetry.

\section{Asymptotic form of $w_N(H_1, H_2)$ for large $N$ and the symmetry breaking}
\label{sec_asy}
In this section, we calculate the asymptotic form of $w_N(H_1, H_2)$ for large $N$. 
Taking the thermodynamic limit before turning the magnetic fields off, 
we show that the symmetry exchanging the copies 1 and 2 is spontaneously broken.  
\subsection{Evaluation of $I_N$}
The asymptotic value of $I_N$ for large $N$  defined in (\ref{def_INJN}) 
 is calculated with use of  properties 
of $f(t)$ clarified in Refs.~\onlinecite{D81,GD89}.  
The result is 
\beq
  I_N \simeq 
\left\{
  \begin{array}{ll}
   1 & ( \be < \be_c )\\
   \frac{\be_c}{\be} & (\be > \be_c), 
  \end{array}
  \right.
  \label{res_IN}
\eeq 
where $\be_c :=2\sqrt{\log 2}/J$ is the critical temperature. 
The calculation deriving (\ref{res_IN}) is lengthy, so that we show it in Appendix. 
In the main text, we derive the same result with help of 
the susceptibility $\chi$ of the REM obtained by Derrida \cite{D81}: 
\beq
  \chi = \lim_{N \to \infty} \frac{\be}{N} \overline{\bra M^2 \ket - \bra M \ket^2} = 
   \left\{
  \begin{array}{ll}
   \be & ( \be < \be_c )\\
   \be_c & (\be > \be_c).
  \end{array}
  \right.
  \label{res_chi}
\eeq 
From (\ref{form_M;M}) and (\ref{def_wN}) we see that $\chi$ is calculated as 
\beq
  \chi = \lim_{N \to \infty} \(\del_1^2 w_N (0, 0) - \del_1 \del_2 w_N (0, 0) \)
  \label{form_chi}
\eeq
in our formulation. 
For sufficiently large $N$, the summations in $A$ and $B$ 
can be evaluated by their extremum.  Namely, 
\beqa
    A &\simeq& \(1 -  I_N \) e^{N(s(m^*) + \be (H_1+H_2) m^* )}, \nn\\
    B &\simeq&  I_N e^{N(s(m_1^*) + \be H_1m_1^* +s(m_2^*) + \be H_2 m_2^* )}. 
\label{form_AB}
\eeqa
In (\ref{form_AB}), 
$s(m)$ is the following asymptotic form of $n(M) 2^{-N}$ with fixed $m:= M/N$: 
\beq
  s(m) := - \frac{1}{2} \(\(1-m\) \log\(1-m\) + \(1+ m\) \log\(1+ m\) \), 
  \label{def_sm}
\eeq
and 
\beqa
  m^* &:=& \tanh\be (H_1+H_2), \nn\\
  m^*_a &:=& \tanh \be H_a \qquad (a=1,2), 
  \label{res_m*}
\eeqa
which are respectively the solutions of the the extremum conditions
\beq
  s'(m^*) + \be (H_1+H_2) =0, \qquad  s'(m_a^*) + \be H_a =0. 
  \label{eq_extremum}
\eeq

Using (\ref{form_AB}), (\ref{res_m*}) and (\ref{eq_extremum}) in (\ref{def_wN}), 
we find the following asymptotic forms of the  first and the second 
derivatives of $w_N(H_1, H_2)$ for large $N$:  
\begin{widetext}
\beqa
  \del_a w_N (H_1, H_2) &\simeq& \frac{m^* A + m^*_a B}{A+B}
  \nn\\
  \del_a^2 w_N (H_1, H_2) &\simeq&  \frac{\be}{A+B} 
  \left( A (1-m^{*2} ) + B \left(1-m_a^{* 2}\right) \right)
  + \frac{N \be AB \left(m^* - m_a^*\right)^2}{(A+B)^2}
  \nn\\
  \del_1 \del_2 w_N(H_1, H_2) &\simeq&   \frac{\be A \left(1-m^{*2} \right)}{A+B} + 
  \frac{N \be (m^*-m_1^*)(m^*-m_2^*)AB}{(A+B)^2}, 
  \label{form_hesswN}
\eeqa
\end{widetext}
where $a = 1,2$. 
Since $A = 1-I_N, B=I_N, m^*=m^*_1 = m^*_2 = 0$ when $H_1=H_2=0$, we can readily 
calculate (\ref{form_chi}) as  
\beq
  \chi =\lim_{N \to \infty} \(( \be - \be(1-I_N) \) = \be \lim_{N \to \infty} I_N. 
  \label{res_chi2}
\eeq
Comparing (\ref{res_chi2}) with the known result  (\ref{res_chi}), 
we obtain (\ref{res_IN}).  

\subsection{RSB-like Symmetry breaking}
Now we derive the asymptotic form of the Hessian matrix defined
 by (\ref{def_hesswN}). 
In the high-temperature phase, 
$\be < \be_c$,  we find from (\ref{res_IN}) and (\ref{form_AB}) 
that $A$ vanishes in (\ref{form_hesswN}).  Thus, we get
\beq
\be^{-1} w_N^{(2)}(H_1, H_2) \simeq 
  \left(
  \begin{array}{cc}
  1-m_1^{*2}& 0 \\ 
  0 & 1-m_2^{*2}
  \end{array}
  \right)
  \label{res_wN2-0}
\eeq
for sufficiently large $N$. 

In the low-temperature phase, $\be > \be_c$, 
we compare the exponent of $A$ and $B$ introducing 
\beqa
  g(H_1, H_2) &:=& s(m^*) + \be \( H_1+H_2 \) m^* 
  \nn\\
 && -  \sum_{a=1}^2 \left( s(m_a^*) + \be  H_a m_a^* \right). 
\eeqa
We see from (\ref{form_AB}) that $g(H_1, H_2) \simeq (\log A - \log B )/N$. 
For fixed $H_2 > 0$, we can show that the function of $H$, $g(H, H_2)$, is 
monotone increasing and 
$g(0, H_2) = 0$. It means that 
$A$ exponentially dominates over $B$ when $H_1>0$ and $H_2 >0$. 
On the other hand, $B$ exponentially dominates over $A$ 
when $H_1 < 0$ and $H_2 >0$.  Similar calculation leads to 
\beq
\begin{array}{ll}
  \frac{1}{N} \log A > \frac{1}{N} \log B & (H_1 H_2 > 0) \\[2mm]
   \frac{1}{N} \log A < \frac{1}{N} \log B & (H_1 H_2 < 0) \\[2mm]
    \frac{1}{N} \log A \simeq \frac{1}{N} \log B & (H_1 H_2 = 0)
  \end{array}
  \label{res_AB}
\eeq
for large $N$.  It means that, when $H_1 H_2 >0$, 
\beq
  \frac{A}{A+B} \to 1, \ \ \frac{B}{A+B} \to 0
\eeq
in (\ref{form_hesswN}) for example. 
Applying  similar formulas, we get

(i) for $H_1 H_2 >0$, 
\beq
\be^{-1} w_N^{(2)}(H_1, H_2) \simeq 
  \left(
  \begin{array}{cc}
  1-m^{*2} & 1-m^{*2}  \\ 
 1-m^{*2} & 1-m^{*2} 
  \end{array}
  \right)
  \label{res_wN2-1}
\eeq

(ii) for $H_1 H_2 < 0$, 
\beq
\be^{-1} w_N^{(2)}(H_1, H_2) \simeq 
  \left(
 \begin{array}{cc}
  1-m_1^{*2} & 0  \\ 
  0  & 1-m_2^{*2}
  \end{array}
  \right)
    \label{res_wN2-2}
\eeq

\begin{widetext}
(iii) for $H_1 = 0$, 
\beqa
\be^{-1} w_N^{(2)}(0, H_2) \simeq 
  \left(
 \begin{array}{cc}
 1-(1-I_N)\, m_2^{*2} + N I_N(1-I_N) \,  m_2^{*2} & \ (1-I_N)(1-m_2^{*2})  \\ 
  (1-I_N)(1-m_2^{*2})   & 1-m_2^{*2}
  \end{array}
  \right)
    \label{res_wN2-3}
\eeqa

(iv) for $H_2 = 0$, 
\beq
\be^{-1} w_N^{(2)}(H_1, 0) \simeq 
  \left(
 \begin{array}{cc}
1-m_1^{*2}  & (1-I_N)(1-m_1^{*2})  \\ 
  (1-I_N)(1-m_1^{*2}) \ & \ 1-(1-I_N)\, m_1^{*2} + N I_N(1-I_N) \,  m_1^{*2}
  \end{array}
  \right). 
    \label{res_wN2-4}
\eeq
\end{widetext}
Now we consider the symmetry transformation
exchanging the two copies 1 and 2 coupled with $H_1$ and
$H_2$ respectively. For finite $N$,  the symmetry ensures that 
$w_N^{(2)}(0,0)$ is a replica symmetric matrix  
as we have seen in (\ref{form_finiteN}). 
To see the spontaneous symmetry breaking, we first put the symmetry breaking field 
$(H_1, H_2) = (H, 0)$, then take the 
thermodynamic limit $N \to \infty$, and finally turn off $H$. 
In the high-temperature phase, using (\ref{res_wN2-0}), we have 
\beq
  \lim_{H \to 0} \lim_{N \to \infty} w^{(2)}_N(H, 0) 
  = 
  \left(
 \begin{array}{cc}
 \be & 0  \\ 
  0  & \be
  \end{array}
  \right). 
  \label{res_s}
\eeq
It implies  that the symmetry exchanging the copies
 holds in the high-temperature phase.  On the other hand, in the 
low-temperature phase, it is found from (\ref{res_wN2-4}) that  
\beq
  \lim_{H \to 0} \lim_{N \to \infty} w^{(2)}_N(H, 0) 
  = 
  \left(
 \begin{array}{cc}
 \be & \be - \be_c \\ 
  \be - \be_c & + \infty
  \end{array}
  \right). 
  \label{res_ssb}
\eeq
 It is no longer a replica symmetric matrix, so that the symmetry exchanging the two copies
  is spontaneously broken in the usual sense of the statistical machenics.  
The symmetry breaking is reminiscent of the RSB in which the zero-replica 
limit not mathematically justified is inevitable \cite{MM09}. 
The symmetry breaking presented here 
can be a mathematically well-defined counterpart to the RSB. 

\begin{figure}[h]
\begin{center}
\setlength{\unitlength}{1mm}
%\begin{picture}(100, 40)(0,0)
 %       \put(0,0){ 
		\includegraphics[scale=0.6, bb=114 326 477 527]{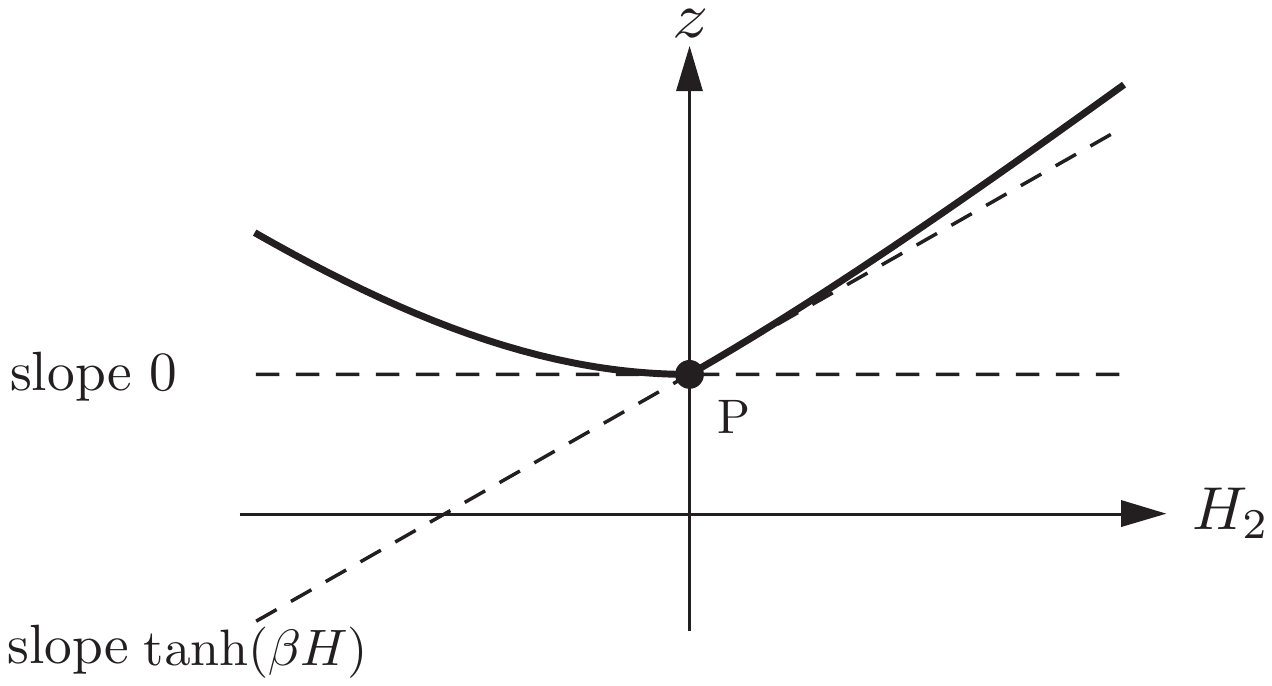}
%		}
% \end{picture}
\end{center}
\caption{
The solid curve is $z = w(H, H_2)$ with fixed $H>0$. 
The dashed lines represent tangential lines at P$(H, 0, w(H, 0))$ 
in the $H_1H_2z$ space. 
The slopes $0$ and $\tanh(\be H)$
are respectively the left and the right derivative at $H_2=0$, which 
correspond to the spontaneous magnetization in the copy 2 
in  presence of the magnetic field $H$ in the copy 1. }
\label{fig_wSection1}
\end{figure}

It is worthwhile exploring the surface defined by $z=w_N(H_1, H_2)$ in the thermodynamic limit 
for understanding the broken symmetry.  Let us define 
\beq
  w(H_1, H_2) := \lim_{N \to \infty} w_N(H_1, H_2) = 
  \lim_{N \to \infty}\frac{1}{N\be} \log(A+B). 
  \label{def_w}
\eeq
In the high-temperature phase, $A$ vanishes in (\ref{def_w}) 
since $I_N \to 1$ as $N \to \infty$ according to 
(\ref{res_IN}). We immediately find from (\ref{form_AB}) that 
\beq
  w(H_1, H_2) = \sum_{a=1}^2 \left( \frac{s(m_a^*)}{\be} + H_a m_a^* \right)
  \label{form_highw}
\eeq
for all $H_1$ and $H_2$.  It is analytic and
derives (\ref{res_s}). 
Using the relationship (\ref{res_AB}),  we can derive $w(H_1, H_2)$ 
in the low-temperature phase in a similar manner.  The result is 
\begin{widetext}
\beqa
  w(H_1, H_2) = \left\{
  \begin{array}{ll}
   \frac{1}{\be} s(m^*) +  (H_1 + H_2) m^*  & (H_1 H_2 \geq 0)\\[2mm]
  \sum_{a=1}^2 \left( \frac{1}{\be} s(m_a^*) +  H_a m_a^* \right) & (H_1 H_2 <0)\\[2mm]
  \end{array}
  \right..
  \label{form_loww}
\eeqa
\end{widetext}
It is continuous on the whole $H_1H_2$ plane. 
When $H_1 H_2 \neq 0$,  $w(H_1, H_2)$ is differentiable and 
\beq
   \del_a w(H_1, H_2) = \left \{
  \begin{array}{ll}
  m^{*}  & (H_1 H_2 > 0) \\
  m_a^{*} & (H_1 H_2 <0)
  \end{array}
  \right.
  \ \ (a =1, 2). 
  \label{form_phi}
\eeq
It shows that 
$\del_a w(H_1, H_2)$ $(a=1,2)$ is discontinuous on $H_a=0$ except the origin. 
For example, when $H>0$, we get
\beqa
  \lim_{H_2 \uparrow 0} \del_2 w(H,H_2) &=& \lim_{H_2 \uparrow 0} m_2^* = 0,  
  \nn\\ 
 \lim_{H_2 \downarrow 0} \del_2 w(H, H_2) &=& 
 \left. m^* \right|_{H_1=H, H_2=0} 
 \nn\\
 &=& \tanh\(\be H\) \neq 0. 
\label{form_del2w}
\eeqa 
This non-differentiability, which is depicted in Fig.\ref{fig_wSection1}, 
  leads to $\del_2^2 w(H, 0) = + \infty$ in (\ref{res_ssb}).  On the other hand, 
  since $w(H, 0)$ is infinitely many-times 
  differentiable with respect to $H$, $\del_1^2w(H, 0) < \infty $.  
It indicates that the diagonal part of the Hessian differs each other at $(H_1, H_2)=(H, 0)$ 
for an arbitrary $H\neq 0$, which results in the spontaneously symmetry breaking.
%
%
%
%%%Figure%%%
%%%end of Figure%%%
%
%

A physical picture of this non-differentiability will be  explained as follows:
first we set $H_1=H_2=0$. In order to obtain a magnetization in the copy 1, 
we need to put a finite external field $H$ to the copy 1, which yields the magnetization 
$m_1 = \tanh (\be H)$. It means that the copy 1 shows  
paramagnetism. Next, we put an \textit{infinitesimal} magnetic field $H_2$ 
to the copy 2. If $H_2$ has the same direction as $H$ ($H H_2 >0$), the 
copy 2 has the spontaneous magnetization with just the same value as $m_1$. 
On the other hand, if $H_2$ has the opposite direction as $H$,  the copy 2 
has no longer finite magnetization.  In this way,  a value of the spontaneous magnetization 
is different depending on a direction of the infinitesimal magnetic field, 
which results in the non-differentiability.  This picture implies the meaning of 
the broken symmetry exchanging 1 and 2. Namely, if we want to magnetize both 
the copy 1 and the copy 2, we need to apply a finite external field to one copy, while 
 it is sufficient to apply an infinitesimal field to the other copy.

\section{The Effective Potential}
\label{sec_eff}
In this section,  we derive
 the effective potential $\ga(\varphi_1, \varphi_2)$ 
 conjugate to $w(H_1, H_2)$, which is defined by the 
 following Legendre transformation
 \beq
   \ga (\varphi_1, \varphi_2) := 
   \max_{H_1, H_2} \( \varphi_1 H_1 + \varphi_2 H_2 - w\(H_1, H_2\) \). 
   \label{def_ga}
 \eeq
Here,  if $w(H_1, H_2)$ is differentiable, the maximization can be carried out 
by solving the following equations 
\beq
  \varphi_a = \del_a w(H_1, H_2), \ \ (a = 1,2 ) 
  \label{form_varphi}
\eeq
for $H_1$ and $H_2$, and then inserting the solutions into the right-hand side of (\ref{def_ga}). 

 In the high-temperature phase,  $w(H_1, H_2)$ is given by 
 the formula (\ref{form_highw}). Using the extremum condition (\ref{eq_extremum}),
we get 
\beq
  \varphi_a  = m_a^*,  \ \ (a = 1,2 ) 
\eeq
for all $H_1$ and $H_2$. Solving them for $H_1$ and $H_2$, we  obtain 
\beq
  \ga(\varphi_1, \varphi_2) = - \frac{1}{\be} \sum_{a=1}^2 s(\varphi_a). 
\eeq
It has the global minimum at the origin and has no singularity. 

In the low-temperature phase,  using (\ref{form_loww}) and (\ref{form_phi}), 
we can derive $\gamma(\varphi_1, \varphi_2)$ 
in a similar manner.  We have 
\beq
  \ga(\varphi_1, \varphi_2) = \left \{
  \begin{array}{ll}
 - \frac{1}{\be}  s(\varphi_a),    & (\varphi_1 = \varphi_2) \\
  - \frac{1}{\be} \sum_{a=1}^2 s(\varphi_a)  & (\varphi_1 \varphi_2 <0)
  \end{array}
  \right..
  \label{form_ga}
\eeq
In the above formula, note that the domain defined by $H_1 H_2 >0$ maps to the line 
$\varphi_1 = \varphi_2$. 
In order to determine  $\ga(\varphi_1, \varphi_2)$ for all $\varphi_1$ and $\varphi_2$
($|\varphi_a| < 1, \, a=1,2)$, 
we have to investigate the 
case of $H_1 H_2=0$.  In this case, a partial derivative does not exist as we have seen in 
the previous section, so that we employ 
  the following geometrical meaning of the Legendre transformation (\ref{def_ga}):
for a given $\varphi_1$ and $\varphi_2$,  consider the plane defined by the following formula 
\beq
 z = \varphi_1 H_1 + \varphi_2 H_2 +z_0
 \label{def_plane}
 \eeq
in the $H_1 H_2 z$ space. 
We choose $z_0$ in such a way that the plane has a common point with 
the surface $z=w(H_1, H_2)$ and try to minimize the value of $z_0$.  
The minimum value of $z_0$ gives 
$-\ga(\varphi_1, \varphi_2)$.

\begin{figure}[h]
\begin{center}
\setlength{\unitlength}{1mm}
%\begin{picture}(100, 40)(0,0)
%        \put(0,0){ 
		\includegraphics[scale=0.6, bb=182 340 464 522]{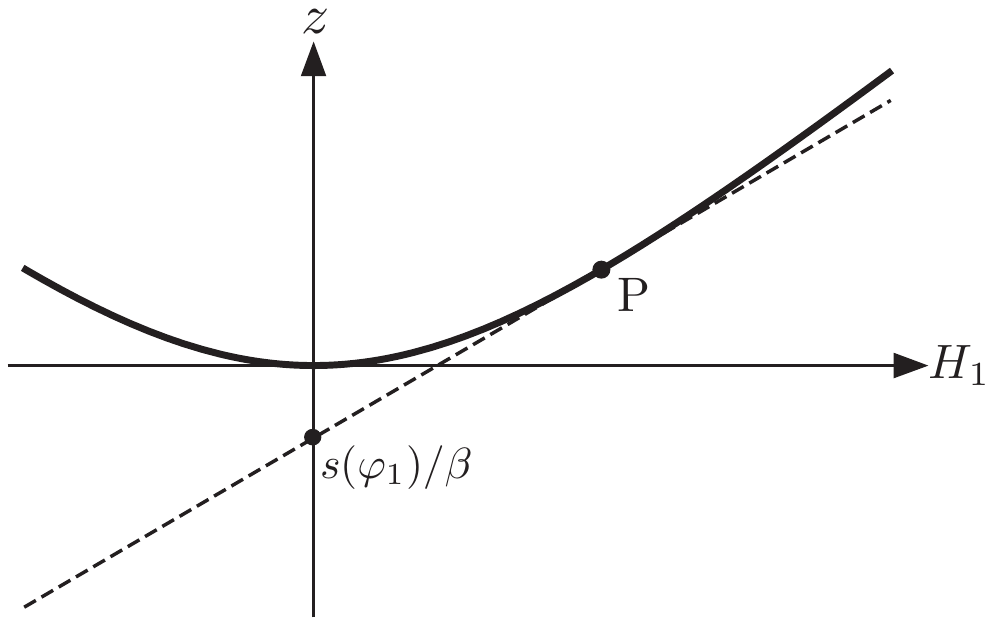}
%		}
% \end{picture}
\end{center}
\caption{The sectional plane $H_2 =0$ in the $H_1H_2z$ space.  
The solid line is the cross section of the surface $z = w(H_1, H_2)$. 
The dashed line represents the the plane $z=\varphi_1 H_1 + \varphi_2 H_2 +z_0$ contacting with the surface at $\textrm{P}(H, 0,  w(H, 0))$. 
It intercepts the $z$ axes at 
$s(\varphi_1)/\be$, which is equal to $-\ga(\varphi_1, \varphi_2)$. }
\label{fig_wSection2}
\end{figure}

First we consider the case of $H_2=0$.  Take an arbitrary point $(H, 0)$ on the line 
and consider the corresponding point $\textrm{P}(0, H, w(H, 0))$ on the surface 
$z = w(H_1, H_2)$.  
Choosing $\varphi_1$, $\varphi_2$ and $z_0$ in (\ref{def_plane}) appropriately, 
we construct a plane 
contacting with the surface $z=w(H_1, H_2)$ at P.  
Since $\del_1 w(H, 0)$ is well-defined  according to (\ref{form_loww}), 
$\varphi_1$ is uniquely determined as 
\beq
\varphi_1 = \del_1 w(H, 0) = \left. m^*\right|_{H_1=H, H_2=0} = \tanh(\be H). 
\eeq
On the other hand,  $\del_2 w(H, 0)$ does not exist as we have seen in (\ref{form_del2w}). 
In this case, $\varphi_2$ can take the value 
 between the left and the right derivative, namely, 0 and 
$\tanh(\be H) = \varphi_1$. 
Since  the point P is on the plane (\ref{def_plane}), 
we find that $z_0 = w(H, 0) - \varphi_1 H = s(\varphi_1) /\be$. 
See Fig \ref{fig_wSection2}.  Consequently  the plane (\ref{def_plane}) contacts with 
$z=w(H_1, H_2)$ at P if  $\varphi_1 = \tanh(\beta H)$, $z_0 = s(\varphi_1) /\be$, 
and $\varphi_2$ is a value  between 0 and $\varphi_1$. 
%
%
%%%Figure%%%
%%%end of Figure%%%
%
%
%%
%

Note that if $z_0$ took a value less than $s(\varphi_1) /\be$, the plane 
(\ref{def_plane}) would not have a common point with the surface. 
Thus we conclude that 
\beq
 \ga\(\varphi_1, \varphi_2 \) = -s(\varphi_1) /\be
 \label{form_ga2}
\eeq
for $\varphi_2 \in [0, \varphi_1]$ or $\varphi_2 \in [\varphi_1, 0]$. 

When $H_1=0$,  exchanging the role of $\varphi_1$ and $\varphi_2$ in 
the case of $H_2=0$, 
we get 
\beq
 \ga\(\varphi_1, \varphi_2 \) = -s(\varphi_2) /\be
 \label{form_ga3}
\eeq
for $\varphi_1\in [0, \varphi_2]$ or $\varphi_1 \in [\varphi_2, 0]$.

\begin{figure}[h]
\begin{center}
\setlength{\unitlength}{1mm}
%\begin{picture}(100, 40)(0,0)
%        \put(0,0){ 
		\includegraphics[scale=0.5, bb=183 162 520 493]{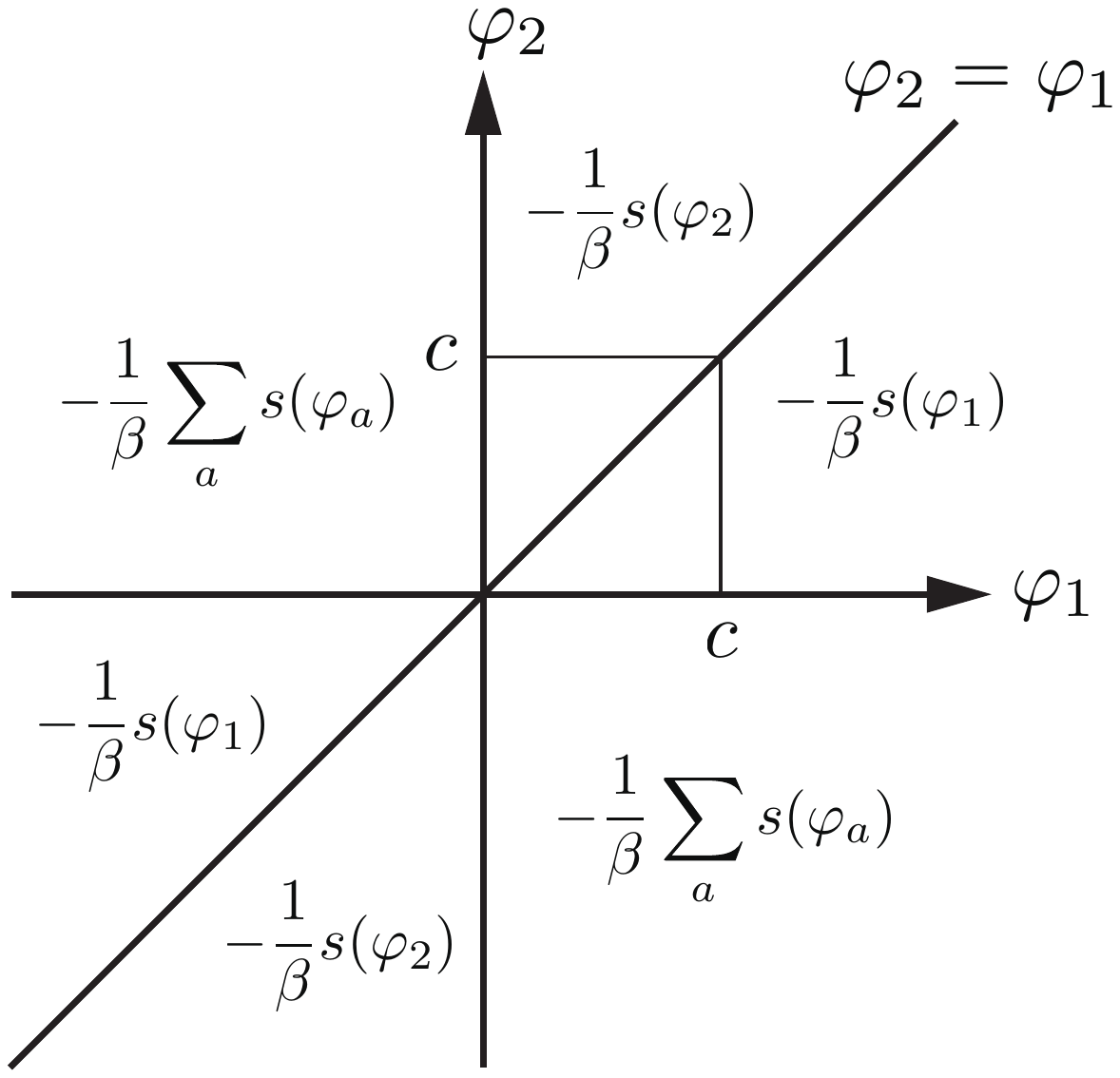}
%		}
% \end{picture}
\end{center}
\caption{Values of $\ga(\varphi_1, \varphi_2)$ on the $\varphi_1 \varphi_2$ plane. 
The segments on $\varphi_1 =c$ or $\varphi_2=c$ show a contour with the 
value $\gamma(\varphi_1, \varphi_2)=-\frac{1}{\beta} s(c)$. They meet at $\varphi_1=\varphi_2=c$, where the effective potential becomes singular. See also Fig. \ref{fig_flatGamma}. }
\label{fig_gamma}
\end{figure}

Combining the results (\ref{form_ga}) (\ref{form_ga2}) and (\ref{form_ga3}), 
we finally obtain
\begin{widetext}
\beq
  \ga(\varphi_1, \varphi_2) = \left \{
  \begin{array}{ll}
 - \frac{1}{\be}  s(\varphi_1)    &
  (0 \leq \varphi_2 \leq \varphi_1 \ \textrm{or} \  \varphi_1 \leq \varphi_2 \leq 0)\\[2mm]
  - \frac{1}{\be}  s(\varphi_2)   &
  (0 \leq \varphi_1 \leq \varphi_2 \ \textrm{or} \  \varphi_2 \leq \varphi_1 \leq 0)\\[2mm] 
  - \frac{1}{\be} \sum_{a=1}^2 s(\varphi_a) & (\varphi_1 \varphi_2 <0)
  \end{array}
  \right..
  \label{res_ga}
\eeq
\end{widetext}
As is shown in Fig.\ref{fig_gamma}, regions that specify the values of 
$\gamma(\varphi_1, \varphi_2)$ have the boudaries $\varphi_a = 0$ 
$(a=1,2)$ and $\varphi_2 = \varphi_1$, on which it is continuous but non-analytic. 
%
%
%%%Figure%%%
%%%end of Figure%%%
%
%

The non-analyticity on $\varphi_1 = \varphi_2$ is similar to 
behavior observed in fixed-point potentials of 
the FRG in various disordered systems \cite{F86,Fe01,LW02,TT04,WL07,LMW08,TT08}.  
Because of a property of  disorder correlators in that literature, the potential 
term in a replicated Hamiltonian depends on the variable $|\varphi_1 - \varphi_2|$ 
and has a singularity at $|\varphi_1 - \varphi_2| =0$.  
Hence it is helpful for comparison to introduce the variables $x:=(\varphi_2 - \varphi_1)/2$ and 
$y:= (\varphi_2 + \varphi_1)/2$.  For fixed $y >0$ and for small $x$ satisfying $|x|<y$, 
the effective potential is written as 
\beq
  \ga(\varphi_1, \varphi_2) = -\frac{1}{\be} s\( y + |x|\). 
\eeq
We see the singularity at $x=0$, which is similar to the fixed-point potential 
in the random $O(N)$ model studied in \cite{LW02}. 
 
\begin{figure}[h]
\begin{center}
\setlength{\unitlength}{1mm}
%\begin{picture}(100, 40)(0,0)
%        \put(0,0){ 
		\includegraphics[scale=0.45, bb=41 410 562 579]{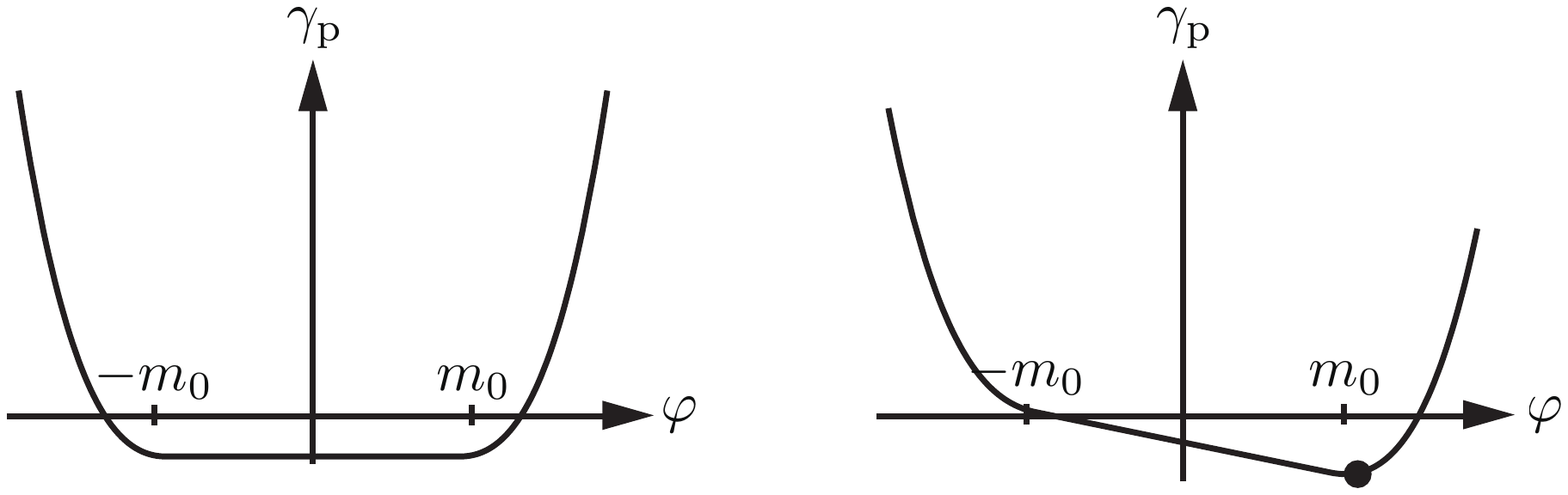}
%		}
% \end{picture}
\end{center}
\caption{A form of the effective potential $\ga_\textrm{p} (\varphi)$ of a 
pure $Z_2$ spin theory (left).  It has a flat bottom $-m_0 \leq \varphi \leq m_0$. 
In order to take a unique minimizer,  we put an infinitesimal 
magnetic field $H$ 
which yields the additional term 
$-\varphi \, H$ to $\ga_\textrm{p}(\varphi)$, thus the shape of the effective potential 
becomes the right figure.  The unique minimizer denoting the dot in the figure 
approaches $m_0$ as $H \to 0$. }
\label{fig_pureZ2}
\end{figure}

Now we consider a physical picture suggested by the singularities including $\varphi_1=0$ 
or $\varphi_2=0$.  
Let us  recall  a form of an 
 effective potential in  $Z_2$ symmetric (pure) spin theory.  
 It is well known that a minimizer of the effective potential gives the value of 
 spontaneous magnetization. 
 In the low-temperature phase, the classical potential has a shape of a double well. 
 However, since the effective potential must be a convex function, 
 it has a flat bottom as depicted in Fig. \ref{fig_pureZ2} \cite{G92}.   %
%
%%%Figure%%%
%%%end of Figure%%%
%
%
In order to take 
 a unique minimizer, we need to turn on an infinitesimal magnetic field. 
 The resultant minimizer is a function of the magnetic field and does not vanish  
 after the magnetic field is turned off. It 
 corresponds to the value of the spontaneous magnetization. 
 
\begin{figure}[h]
\begin{center}
\setlength{\unitlength}{1mm}
%\begin{picture}(100, 40)(0,0)
%        \put(0,0){ 
		\includegraphics[scale=0.5, bb=182 341 478 543]{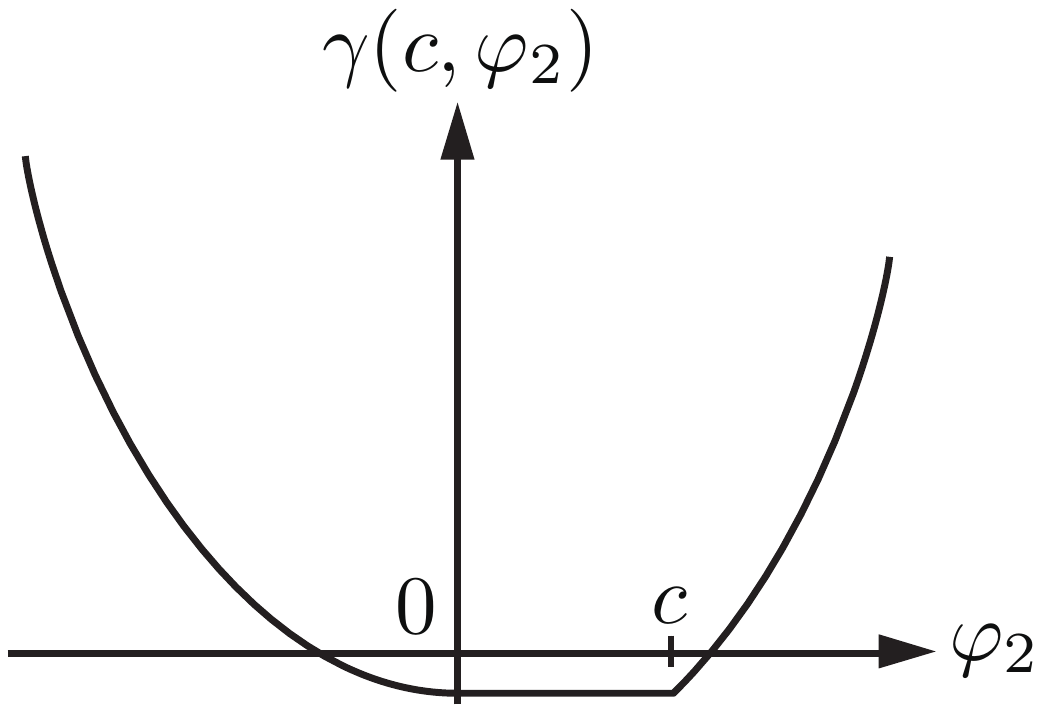}
%		}
% \end{picture}
\end{center}
\caption{A graph of  $\ga (c, \varphi_2)$ as 
a function of $\varphi_2$ in the case when $c>0$.
We see from (\ref{res_ga}) that  it has a flat bottom $0 \leq \varphi_2 \leq c$. 
We turn on an infinitesimal 
magnetic field $H$ for selecting the unique minimizer. 
If $H >0$, the value of the spontaneous magnetization in
the copy 2 is $c$, which is the same value as in the copy 1.  
On the other hand,  if $H <0$, the spontaneous magnetization vanishes.  }
\label{fig_flatGamma}
\end{figure}

Now we apply the idea to $\ga(\varphi_1, \varphi_2)$ in the low-temperature phase. 
It is found from (\ref{res_ga}) that 
$\ga(\varphi_1, \varphi_2)$ has a unique minimum at the origin, which corresponds to the fact
that there is no spontaneous magnetization. 
First a magnetic field is turned on to the copy 1 in such a way that the 
minimizer is shifted to $(\varphi_1, \varphi_2)=(c, 0)$ $(c\neq 0)$.  Next an infinitesimal 
magnetic field having the sign \textit{same}  
as $c$ is turned on in the copy 2.  
As we see from Fig. \ref{fig_flatGamma}, this yields 
 spontaneous magnetization in the copy 2 with the value $c$, 
 just the same value as in the copy 1.  This originates from the non-analyticity 
 on $\varphi_1 = \varphi_2$. 
However, if the infinitesimal field has the \textit{opposite} sign to $c$, 
the value of the magnetization vanishes due to the singularity on $\varphi_2=0$.  
%
%
%
%
%%%Figure%%%
%%%end of Figure%%%
%
%
%This result can be interpreted that each spin in the copy 2 can flip so as to have 
%the same value in the copy 1 with the help of the infinitesimal magnetic field,  
%while a spin configuration in the copy 1 do not effect to the copy 2 at all 
%when the infinitesimal filed disturbs. This feature may characterize glassy phases
%in general. 

\section{Summary and Discussion}
Introducing two copies of the REM coupled with independent magnetic fields,  we have calculated the generating function for the correlator of the magnetization.  
When the system is finite, the Hessian of the generating function 
at the zero-magnetic fields is a replica symmetric matrix.  
It reflects the symmetry exchanging the two copies. 
In the low temperature phase, however, 
we see that the symmetry is spontaneously broken in the usual sense of the statistical mechanics.
In fact, one of the diagonal components of the Hessian
 becomes infinity while the other remains finite. The asymmetry of the diagonal components is physically interpreted as follows: if we want to magnetize the system, 
 we have to turn on external magnetic field to one copy as in the case of the paramagnetism, 
 while we can see spontaneous magnetization in the other copy.  

This broken symmetry reminds us of the usual RSB and may provide a rigorous notion for the RSB. 
In order to clarify this observation, we need to investigate other disordered systems and show the universality of the broken symmetry presented in this work. If it exists in the various disordered systems, one also has to consider the relationship to the usual RSB and to glassy behavior. We are now planning to study the REM having
 the ferromagnetic coupling \cite{MSxx}. 

Furthermore,  the value of the magnetizations coincide each other if the two magnetic fields, 
one is finite and the other is infinitesimal for the spontaneous magnetization, 
have the same direction. Coincidence of the two magnetization reflects the fact that 
the effective potential corresponding  to the generating function has the singularity at which 
the two independent variables coincide. It supports the fact that singularity of the fixed-point 
potential of the FRG certainly exists in disordered systems, not the artifact of approximation. 

\begin{acknowledgments}
The author would like to thank G. Tarjus, V. Dotsenko and M. Tissier  for fruitful discussions.   
He is also grateful to LPTMC (Paris 6) for kind hospitality, where most of this work has been done. 
\end{acknowledgments}
\appendix*
\section{Asymptotic value of $I_N$}
\label{appendix}
In this appendix, we derive the asymptotic value (\ref{res_IN}) evaluating $f(t)$ 
defined by (\ref{def_ft}).   
It  has been first introduced and investigated by Derrida \cite{D81}. 
A similar analysis has been recently performed by Dotsenko \cite{D11} in which 
the same function is called $G(N, x)$. Here, 
we follow the analysis carried out by Gardner and Derrida \cite{GD89}. 

Using (\ref{def_P}), we can write $f(t)$ as 
\beq
  f(t) :=  \overline{e^{-t e^{-\be E_{M, k}}}}  = 
  \frac{1}{\sqrt{\pi}} \int_{-\infty}^{\infty} dy \, e^{-y^2 - t e^{- \la y}}, 
  \label{form_ft0}
\eeq
where $\la := \sqrt{N} \be J$. It has 
 the following asymptotic form depending on ranges of $\log t$ \cite{D81}:
\beq
  f(t) \simeq 
  \left\{
  \begin{array}{ll}
  - k(t) \, e^{-(\log t)^2/\la^2}
  & (\log t >0)
  \\[2mm]
  1 -  k(t) \, e^{-(\log t)^2/\la^2}
  & ( -\la^2/2<\log t  < 0 )
  \\[2mm]
  1 - e^{\la^2/4} t &  ( \log t  < -\la^2/2 )
  \end{array}
  \right., 
  \label{form_ft}
\eeq 
where $k(t)$ is the function of $t$ defined by  
\beq
  k(t) := \frac{-\Gamma\(\frac{2\log t}{\la^2} \)}{\sqrt{\pi}\la}. 
\eeq
Although higher-order terms with respect to $t$ are determined when 
$\log t < -\la^2/2$ in Ref. \onlinecite{D81}, the main terms described in (\ref{form_ft}) are 
sufficient in the present study. 
Introducing $\phi(t)$ by the following formula
\beq
  \exp\(-\phi(t) \) = \( f(t) \)^{2^N}, 
  \label{def_phit}
\eeq
the integral $I_N$ defined in (\ref{def_INJN}) is written as 
\beq
  I_N = 2^{2N}\int_0^\infty d t\, t \, \(\frac{f'(t)}{f(t)} \)^2 e^{-\phi(t)}. 
\eeq
We divide the interval $[0, \infty) \ni t$ into the following three intervals 
\beq
  K_1 := [0, e^{-\la^2/2}], \ \ 
  K_2 := [e^{-\la^2/2}, 1],  \ \ 
  K_3 := [1, \infty)
\eeq
in accordance with  (\ref{form_ft}), and evaluate 
\beq
  I^{(j)} :=  2^{2N} \int_{K_j} dt \, t \(\frac{f'(t)}{f(t)} \)^2 e^{-\phi(t)}, 
  \qquad j = 1,2,3
\eeq
separately. 
To begin with, we compute 
\beq
  I^{(1)} = 2^{2N} \int_0^{e^{-\la^2/2}} dt \, t \(\frac{f'(t)}{f(t)} \)^2 e^{-\phi(t)}, 
\eeq
in which $f(t)$ behaves as
\beq
  f(t) \simeq \exp\(-t \, e^{\la^2/4} \).  
\eeq
This yields 
\beqa
  I^{(1)} =  2^{2N} e^{-\la^2/2} \int_0^{-\la^2/2} dt \, t  \, e^{-t \, 2^N e^{\la^2/4}}. 
\eeqa
Changing the variable $s := t \, 2^N e^{\la^2/4}$, one finds that
\beq
  \lim_{N \to \infty} I^{(1)} = \int_0^\infty ds \, s \, e^{-s} = 1
\eeq
when $\be < \be_c = 2\sqrt{\log 2}/J$. 
On the other hand, when $\be > \be_c$, the interval of the integration contracts to $0$. 
Thus, we conclude that
\beq
 \lim_{N \to \infty}   I^{(1)} = 
   \left\{
   \begin{array}{ll}
     1 & (\be < \be_c)\\
     0 & (\be > \be_c)
   \end{array}
   \right. .
   \label{res_I1}
\eeq

Next, we compute 
\beq
  I^{(2)} = 2^{2N} \int_{e^{-\la^2/2}}^{1} dt \, t \(\frac{f'(t)}{f(t)} \)^2 e^{-\phi(t)}. 
  \label{form_I(2)}
\eeq
In this region, $f(t)$ behaves as 
\beq
  f(t) \simeq \exp\(- k(t) \, e^{-(\log t/\la)^2} \). 
\eeq
From (\ref{def_phit}), we have 
\beq
  \phi(t) = - 2^N \log f(t) \simeq 2^N k(t) e^{-(\log t/\la)^2}. 
  \label{def_phi2}
\eeq
Let $x$ be the solution of $\log \phi(e^x) =0$ with $x<0$.  
Namely $x$ is the negative solution of   
\beq
  \log \phi (e^x)= N \log 2 + \log k(t) - x^2/\la^2 = 0. 
  \label{form_logphi}
\eeq
For large $N$, we can derive the leading term of $x$ as  
\beq
  x = - \la \sqrt{N \log 2} + {\textrm{o}}(N), 
  \label{form_x}
\eeq
where $\textrm{o}(N)$ is a part satisfying $\textrm{o}(N) / N \to 0$ as 
$N \to \infty$. If  $t \in (e^x, 1]$ then $\log \phi(t) > 0$. 
It indicates that $\phi(t)$ 
 becomes exponentially large  in $N$ 
 when $ t \in (e^x, 1]$.   In particular, when $e^x < e^{-\la^2/2}$, 
or equivalently, $\be < \be_c$, $\phi (t)$ becomes exponentially large in 
$N$ for all $ t \in K_2$.  It means that 
\beq
  \lim_{N \to \infty} I^{(2)} = 0 
  \label{res_I2-1}
\eeq
in the high-temperature phase. 

When $\be > \be_c$, dominant contribution to $I^{(2)}$ can come from 
the region where 
$\phi(t) \sim \textrm{O}(1)$ 
or $\phi(t)$ is exponentially small in $N$ \cite{GD89}. 
In order to specify the exponentially small region, where $\log \phi (t) < 0$, 
we define 
\beq
  x_\ep := x - N \ep
  \label{def_xep}
\eeq
for sufficiently small $\ep >0$.  We see from (\ref{form_logphi}) that 
$\phi (t)$ is exponentially small in $N$ if $t \in [e^{-\la^2/2},  e^{x_\ep}]$, while a region for 
$\phi(t) \sim {\textrm O}(1)$ is contained in $[e^{x_\ep}, 1]$. 
We separately deal with the both cases  dividing $I^{(2)}$ as 
\begin{widetext}
\beq
  I^{(2)} = 2^{2N} \int_{e^{-\la^2/2}}^{e^{x_\ep}} dt \, t \left(\frac{f'(t)}{f(t)} \)^2 e^{-\phi(t)}
  + 2^{2N} \int_{e^{x_\ep}}^1 dt \, t \left(\frac{f'(t)}{f(t)} \)^2 e^{-\phi(t)}
  =  I^{(2)}_1 +  I^{(2)}_2. 
\eeq
\end{widetext}
Let us first evaluate $ I^{(2)}_1$. 
Explicit calculation using (\ref{form_ft0}) shows that 
\beq
  f'(t) = - e^{\la^2/4} f\(e^{\la^2/2} \, t \). 
  \label{form_f't}
\eeq
It indicates that, from (\ref{form_ft}),  
\beq
  f'(t) \simeq  e^{\la^2/4} \, k(t \, e^{\la^2/2}) \exp\(-\( \log t + \la^2/2 \)^2/\la^2 \)
\eeq
for $\log t + \la^2/2 >0$. When $\phi(t)$ is exponentially small, 
we see from (\ref{def_phi2}) that $f(t)$ is very close to 1. Hence we can write 
\begin{widetext}
\beq
  \(\frac{f'(t)}{f(t)} \)^2 \simeq \(f'(t)\)^2 \simeq 
  e^{\la^2/2} k\(t e^{\la^2/2} \)^2  \exp\(-2\( \log t + \la^2/2 \)^2/\la^2 \).
\eeq
Making the change of variable $u := \log t$, we get  
\beq
  I^{(2)}_1 = 2^{2N} \int_{-\la^2/2}^{x_\ep} d u \, k\(e^{u+\la^2/2} \)^2 e^{-2 u^2/\la^2}  
  \simeq k\(e^{x_\ep + \la^2/2} \)^2 e^{2N \log 2 -2 x_{\ep}^2/\la^2}
\eeq
\end{widetext}
because the most dominant contribution comes from $u=x_\ep$. 
Using (\ref{form_x}) and (\ref{def_xep}), we see that $N \log 2 < x_\ep^2/\la^2$ 
for sufficiently large $N$, hence 
the right-hand side is exponentially small, so that 
\beq
   I^{(2)}_1 \to 0 , \qquad (N \to \infty). 
  \label{res_I2-2}
\eeq

Next, we evaluate $I^{(2)}_2$.  Since the main contribution comes from 
$t \sim e^x$, we determine explicit form of $\phi(t)$ around $t \sim e^x$. 
For this purpose, introducing the variable $v$ from the relation $t = v e^x$, 
we write $\phi(t)$ in terms of $v$ assuming that $v \sim 1$. Since 
$\log \phi(e^x)=0$, 
\beqa
  \log \phi(t) &\simeq& \log k\(v e^x\) + N \log 2 - \(\frac{\log v + x}{\la} \)^2
  \nn\\
  &\simeq& -\frac{2 x}{\la^2} \log v, 
\eeqa
so that 
\beq
  \phi(t) \simeq v^{-2 x/\la^2}= t^{-2 x/\la^2} e^{2 x^2/\la^2}. 
  \label{form_phi2}
\eeq
We can extrapolate this relation to the whole interval $[e^{-x_\ep}, \, 1]$ because 
the relation do not affect the integral when $t > e^x$.  
Thus the integration is evaluated as  
\beqa
  I^{(2)}_2 &=&  2^{2N} \int_{e^{x_\ep}}^1 dt \, t \(\frac{f'(t)}{f(t)} \)^2e^{-\phi(t)} 
  \nn\\
  &=& - \frac{2 x}{\la^2}\int_{e^{2x(x - x_\ep )/\la^2}}^{e^{2 x^2/\la^2}} d\phi \, \phi \, e^{-\phi}, 
  \label{form_I22}
\eeqa
where we have used the following formula derived from  (\ref{def_phit}):
\beq
  \frac{d \phi(t)}{d t}  = -2^N \frac{f'(t)}{f(t)}. 
\eeq
Since $x_\ep < x < 0$, the interval in (\ref{form_I22}) approaches $[0, \infty)$ as 
$N \uparrow \infty$.  Employing (\ref{form_x}) in (\ref{form_I22}), we get 
\beq
  I^{(2)}_2
  \simeq \frac{\be_c}{\be} \int_0^\infty d\phi \, \phi \, e^{-\phi} = \frac{\be_c}{\be}.
  \label{res_I2-3} 
\eeq
From the results (\ref{res_I2-2}) and 
(\ref{res_I2-3}), we have 
\beq
  I^{(2)} = I^{(2)}_1 + I^{(2)}_2 \simeq \frac{\be_c}{\be}
\eeq
for $\be > \be_c$. 
Combining the result in the high-temperature phase, (\ref{res_I2-1}), we  conclude that 
\beq
 \lim_{N \to \infty}   I^{(2)} = 
   \left\{
   \begin{array}{ll}
     0 & (\be < \be_c)\\
     \be_c/\be & (\be > \be_c)
   \end{array}
   \right. .
   \label{res_I2}
\eeq

Finally, let us evaluate 
 \beqa
   I^{(3)} &=& 2^{2N} \int_1^\infty dt \, t \(\frac{f'(t)}{f(t)} \)^2 e^{-\phi(t)} 
   \nn\\
   &=& 2^{2N} \int_1^\infty dt \, t \(\frac{f'(t)}{f(t)} \)^2 \(f(t)\)^{2^N}.
 \eeqa
From (\ref{form_f't}), 
\beq
  \left| f'(t) \right| = e^{\la^2/4} \left| f(e^{\la^2/2} t) \right| \leq 
  e^{\la^2/4} \left| f(t) \right|
\eeq
since $f(t)$ is monotone degreasing, which yields 
\beq
  I^{(3)} \leq 2^{2N}e^{\la^2/2} \int_1^\infty dt \, t  \, \( f(t) \)^{2^N}. 
\label{form_I3}
\eeq
Since the asymptotic form (\ref{form_ft}) is  singular at $t=1$ 
due to the gamma function, we rewrite $f(t)$ in the following way: 
making the change of variable $u:= t e^{-\la y}$, we get 
\beq
  f(t) = \frac{e^{-\frac{\la^2  z^2}{4}}}{\sqrt{\pi} \, \la} \int_0^\infty \frac{du}{u} \, u^z \, 
  e^{-u - (\log u/\la)^2}, 
\eeq
where $z:= 2 \log t/\la^2$. 
Now we divide the interval $[0, \infty) \ni u$ into $[0, 1]$ and $[1, \infty)$, 
and call the corresponding integrals $J_1$ and $J_2$ respectively. 
First we evaluate 
\beq
  J_1 := \frac{e^{-\frac{\la^2  z^2}{4}}}{\sqrt{\pi} \, \la}  \int_0^1 \frac{du}{u} \, u^z \, 
  e^{-u - (\log u/\la)^2}. 
  %\leq \frac{e^{-(\la \, z /2)^2}}{2}. 
  \label{res_J1}
\eeq
Since $u \in [0,1]$ and $z>0$, we find that 
\beq
  u^z \, e^{-u - (\log u/\la)^2} \leq  \, e^{- (\log u/\la)^2}, 
\eeq
which results in 
\beq
  J_1 \leq \frac{e^{-\frac{\la^2  z^2}{4}}}{\sqrt{\pi} \, \la} 
  \int_{-\infty}^0 d(\log u) \, e^{- (\log u/\la)^2} = 
  \frac{e^{-\frac{\la^2  z^2}{4}}}{2}. 
\eeq
Next, we consider 
\beq
  J_2 := \frac{e^{-\frac{\la^2  z^2}{4}}}{\sqrt{\pi} \, \la}  \int_1^\infty \frac{du}{u} \, u^z \, 
  e^{-u - (\log u/\la)^2}.  
  %\leq \frac{e^{-\frac{\la^2  z^2}{4}}}{\sqrt{\pi} \la} \Gamma(z+1). 
\eeq
When $u \in [1, \infty)$,  it is easily seen that 
\beq
  u^z \, e^{-u - (\log u/\la)^2} \leq u^{z+1}  e^{-u}, 
\eeq
which leads to  
\beq
  J_2 \leq \frac{e^{-\frac{\la^2  z^2}{4}}}{\sqrt{\pi} \, \la} 
  \int_1^\infty \frac{du}{u} u^{z+1}  e^{-u} 
  \leq \frac{e^{-\frac{\la^2  z^2}{4}}}{\sqrt{\pi} \, \la} \Gamma(z+1). 
    \label{res_J2}
\eeq
Combining (\ref{res_J1}) and (\ref{res_J2}), we get 
\beq
  f(t) = J_1 + J_2 \leq 
  e^{-\frac{\la^2  z^2}{4}}
  \(\frac{1}{2} + \frac{1}{\sqrt{\pi} \, \la} \Gamma(z+1) \). 
  \label{est_ft}
\eeq

For more convenient form, we use the inequality 
\beq
  \Gamma(z+1) \leq e^{z^2}
  \label{res_gaEst}
\eeq
for $z \geq 0$.  It can be shown by the following 
immediate consequence from the theorem 1 in \cite{VV02}: 
\beq
  \frac{1}{z} \log \Gamma (z+1) -\log(z+1) + 1 < 1 -\gamma
\eeq
for $z > 0$, where $\ga$ is the Euler-Mascheroni constant. Thus we have 
\beq
  \log \Gamma (z+1) \leq z \log(z+1) -\ga z \leq z^2
\eeq
for $z \geq 0$. 

Using (\ref{res_gaEst}) in (\ref{est_ft}),  we get 
\beq
  f(t) \leq e^{z^2-\frac{\la^2  z^2}{4}} \(\frac{1}{2} + \frac{1}{\sqrt{\pi} \, \la}\). 
\eeq
Applying this to (\ref{form_I3}), we get 
\begin{widetext}
\beq
  I^{(3)} \leq
 \frac{2^{2N}\la^2 e^{\frac{\la^2}{2}} }{2} \int_0^\infty d z \, 
  e^{\la^2 z - 2^N (\frac{\la^2}{4}-1)z^2} \(\frac{1}{2} + \frac{1}{\sqrt{\pi} \, \la}\)^{2^N}, 
\eeq
\end{widetext}
where we have changed the integration variable from $t$ to $z$. 
Even though  we extend the interval of $z$ from $[0 , \infty)$ to $(-\infty, \infty)$, 
the inequality is maintained. 
Then the integration is explicitly performed for sufficiently large $N$. The result is 
\beq
  I^{(3)} \leq 
  2^{2N}\la^2 e^{\frac{\la^2}{2}} 
  \sqrt{\frac{\pi}{2^N (\la^2 -4)}} e^{\frac{\la^4}{2^N(\la^2-4)}}
  \(\frac{1}{2} + \frac{1}{\sqrt{\pi} \, \la}\)^{2^N}. 
\eeq
Because the last factor is rapidly decreasing, it turns out that 
\beq
  \lim_{N \to \infty} I^{(3)} = 0. 
  \label{res_I3}
\eeq
From (\ref{res_I1}), (\ref{res_I2}) and (\ref{res_I3}), 
we obtain (\ref{res_IN}). 

\end{document}